\documentclass[twocolumn,prb,aps,amsmath,amssymb,floatfix,showpacs]{revtex4-1}

\usepackage{color}
\usepackage{graphicx}
\usepackage{dcolumn}
\usepackage{bm}

\begin{document}


\title{Projector augmented wave calculation of x-ray absorption spectra at the L$_{2,3}$ edges}

\author{Oana Bun\u au}
 \email{bunau@unizar.es}
\affiliation{IMPMC, CNRS/UPMC, 4 place Jussieu, 75252 Paris, France\\ Universidad de Zaragoza, CSIC, ICMA  y Dept. Fis. Mat. Condensada E-50009 Zaragoza, Spain}

\author{Matteo Calandra}
\affiliation{IMPMC, CNRS/UPMC, 4 place Jussieu, 75252 Paris, France}

\date{\today}

\begin{abstract}
We develop a technique based on density functional theory and the
projector augmented wave method in order to obtain the x-ray absorption cross
section at a general
edge, both in the electric dipole and quadrupole approximations. The method is
a generalization of Taillefumier {\it et al.}. \cite{Taillefumier2002}  
We apply the method to the calculation of the Cu  L$_{2,3}$ edges in fcc 
copper and cuprite (Cu$_2$O), and to the S  L$_{2,3}$ edges in 
molybdenite (2H-MoS$_2$). The role of core-hole effects, modeled in a supercell approach, as well as the
decomposition of the spectrum into different angular momentum channels are studied in detail.  In copper we find that the best agreement with experimental data is obtained when core-hole effects are neglected. 
On the contrary, core-hole effects need to be included both in Cu$_2$O and 2H-MoS$_2$.
Finally we show that a non-negligible component of S L$_{2,3}$ edges in 2H-MoS$_2$ involves transition 
to states with $s$ character at all energy scales. The inclusion
of this angular momentum channel is mandatory to correctly describe the angular dependence of 
the measured spectra. We believe that transitions to $s$ character states are quantitatively significant at the L$_{2,3}$ edges of third row elements from Al to Ar.
\end{abstract}

\maketitle


\section{Introduction}

X-ray absorption near edges structure (XANES) spectroscopy is a very  powerful tool for
condensed matter studies. Its chemical and spatial selectivity
 allow solving crystallographic structures and investigating the
properties of materials, such as magnetism or orbital hybridization.
Unambiguous interpretation of x-ray spectra requires theoretical
insight, in order to determine the origin of spectral features and to understand the effect of the core-hole on the measured spectrum. 

Several theoretical approaches, either in real space\cite{FEFF,FDMNES} or based on periodic boundary conditions,\cite{SPRKKR,Wien2k,Gougoussis2009,CASTEP,Tamura2012} were proven successful to describe XANES spectra in molecules and solids at the 
density functional theory (DFT) level. All-electron based implementations,\cite{Wien2k} the most accurate within DFT, suffer
from significantly larger computational costs than those required by pseudopotential-based methods.\cite{Gougoussis2009,CASTEP,Tamura2012}
In this respect, the advantage of  pseudopotential-based methods is that they allow to tackle larger systems, e.g. composed of several hundreds of atoms. In pseudopotential-based methods 
the atomic core-potential is replaced by a fictitious one,  smoother and less singular, hence requiring a smaller representation in terms of plane waves and consequently less computation time. 
As a consequence, in order to describe properties that depend on core-states, like XANES, one needs to reconstruct the all-electron wavefunction
of the empty valence states.
This is achieved in the framework of the projector augmented wave (PAW) method.\cite{Blochl1994} 

The PAW approach for the calculation of x-ray absorption spectra has been implemented in some freely available  (for K-edge only)\cite{Gougoussis2009} first-principles codes, commercial code distributions\cite{CASTEP,Pickard1997} 
and in other non-distributed software. \cite{Tamura2012,Prendergast2006,Hetenyi2004}
The PAW method has been proven successful in interpreting experimental data \cite{Cabaret2007,Gaudry2007,Juhin2008,Gougoussis2010,Juhin2010,Ilakovac2012,Tamura2012,Prendergast2006,Nolan2006,Bordage2012,Cabaret2013}, thanks to its ability to perform the spectral analysis of the calculated XANES.
Most of the applications of the existing methods involve K-edge calculations, where usually core-hole is reliably modeled in a supercell approach and electronic excitations are well described by DFT.

PAW studies at L$_{2,3}$ edges are not numerous.\cite{CASTEP,Lazar2004,Bunau2012a} One of the main reasons is that, contrary to K edges, many-electron effects are common at L$_{2,3}$ edges. For instance, in the case of L$_{2,3}$ edges of transition elements and M$_{4,5}$ edges of rare earths, excitations cannot no longer be treated at the DFT level, due to the breakdown of the single-particle picture.\cite{Zaanen1985} Consequently, more sophisticated methods based on time dependent DFT,\cite{Schwitalla1998,Bunau2012} multichannel multiple scattering theory\cite{Natoli1990,Kruger2004} or the Bethe-Salpeter equation\cite{Laskowski2010} need to be employed. However, none of the aforementioned first principles methods can rigorously deal with open shells. In this particular case, the ligand field multiplet theory\cite{Thole1985} (LFM) can successfully interpret the spectra, albeit not from first principles. 

When the states probed by the spectroscopy are delocalized - K and L$_1$ edges in general (with the exception of very light materials), L$_{2,3}$ and M$_{2,3}$ edges of elements belonging to the 4$d$ and 5$d$ series -  x-ray absorption spectra can usually be interpreted in the DFT framework. In these cases, the final states, mainly of $p$ (K and L$_1$) or $d$ character (L$_{2,3}$ and M$_{2,3}$ edges) show limited correlation effects and, as a consequence, the structures in a XANES spectrum can be associated to the Kohn-Sham states. 

In this work we focus on the description of L$_{2,3}$ edges within DFT and PAW frameworks. 
The previous work in Ref. \onlinecite{Taillefumier2002} is generalized to the case of a general
edge, for a spin-orbit split core-level.   
In a first step, we validate our method by calculating Cu L$_{2,3}$ edges of metallic Cu. We study the convergence of the spectra with respect to the number of projectors considered in the reconstruction step, i.e. the completeness of the projector basis. It is generally agreed\cite{Taillefumier2002,Gougoussis2009} that at K edges two linearly independent projectors \textit{per} angular momentum channel are sufficient as to insure an accurate reconstruction of the 1$s$ state. Albeit a pre-requisite for PAW applications, other edges, and  L$_{2,3}$ in particular, have not benefit from similar studies.  

Next, we switch to the more challenging cases of the Cu  L$_{2,3}$ edges in cuprite (Cu$_2$O) and S L$_{2,3}$ edges in molybdenite (2H-MoS$_2$).
For both compounds, we perform a detailed analysis of the electronic origin of the spectral features and we assess the effect of the 2$p$ core.
 Concerning Cu$_2$O, we investigate how the empty states probed by the spectroscopy are influenced by the inclusion of the Hubbard $U$ term.  In the case of the hexagonal 2H-MoS$_2$ we interpret the effect of the polarization of the beam, i.e. the x-ray natural linear dichroism (XNLD), by decomposing the spectra into various angular moment channels. 

The article is organized as follows. 
In section \ref{sec:theory} we recall formal aspects of the x-ray
cross section calculation in the PAW formalism.  
In section \ref{sec:calc} we give the technical details of the
calculations.
Next, in section \ref{sec:res}, we present results for the
three aforementioned compounds. 
Finally, in the appendix we include the full analytical expression of the electric dipole (E1) and electric
quadrupole (E2) matrix elements for a general edge.

\section{X ray absorption cross section}
\label{sec:theory}
In the single particle approximation, the x-ray absorption cross section reads:\cite{Cowan}
\begin{equation}
\sigma(\omega)=4\pi^2\alpha \hbar\omega\sum_{f, i}|\langle \psi_f | \hat {\cal O}| \psi_i\rangle|^{2}\,\,\delta\left(\hbar\omega-E_f+E_{i}\right)
\label{eq:sig_onebody}
\end{equation}
where $\psi_i$ is the one-particle wavefunction of the core level and $\psi_f$ is the all-electron one-particle (empty) final state. The corresponding electronic  energies are $E_{i}$ and $E_f$, $\hbar\omega$ being the photon energy and $\alpha$ is the fine structure constant. 
The transition operator is taken as: 
$$ \hat {\cal O} = \hat {\cal D} + \hat {\cal Q} \label{eq:multipoles}$$
where  $\hat {\cal D} = \hat{\bm{\varepsilon}} \cdot \bm r $ and ${\cal Q} = \frac{i}{2}(\hat{\bm{\varepsilon}} \cdot \bm r) (\bm k \cdot \bm r)$ 
are the electric dipole and quadrupole operators, respectively. 
The quantities,
$\hat{\bm{\varepsilon}}$ and $\bm k$ are 
the polarization vector and the wave-vector of the incident x-ray beam, respectively, and  ${\bm r}$ is the position coordinate
of the electron. In this work we only deal with the pure dipolar absorption term (E1E1). The quadrupolar contribution (E2E2) is generally negligible at the L$_{2,3}$ edges of transition elements, but may become important in rare-earths compounds. E1E2 cross-terms appear in optically active materials. 

In the next paragraphs we briefly introduce the PAW method applied to the calculation of the x-ray absorption cross section. We closely follow Refs. \onlinecite{Taillefumier2002} and \onlinecite{Gougoussis2009}.

In a pseudopotential calculation  the pseudo-wavefunction of the crystal 
$| \tilde{\psi}_f \rangle$ is obtained at the end of the self-consistent field run. 
In order to get the all-electron wavefunctions 
$\left| \psi_f \right\rangle$ needed in Eq. \ref{eq:sig_onebody} all electron reconstruction\cite{Blochl1994} needs to be performed. As shown in Refs. \onlinecite{Taillefumier2002}, \onlinecite{Gougoussis2009} and \onlinecite{PARATEC} the cross section may be written as: 
\begin{equation}
\label{eq:pawcrosssec2}
\sigma(\omega) = 4 \pi^2 \alpha \hbar \omega \sum_{f, i} \ \left|\left\langle \tilde\psi_{f} \middle| \tilde\phi_{\mathbf{R}_0}\right\rangle \right|^2 \delta(E_f-E_i-\hbar \omega)
\end{equation}
where 
\begin{equation}
\label{eq:pawphi0}
\left| \tilde\phi_{\mathbf{R}_0}\right\rangle= 
\sum_p \left| \tilde{p}_{\mathbf{R}_0,p} \right\rangle  
\left\langle \phi_{\mathbf{R}_0,p}\middle| \hat{ {\cal O}}\middle| \psi_i \right\rangle \,.
\end{equation}
In Eq. \ref{eq:pawphi0}, $\left|\phi_{\mathbf{R_0},p}\right\rangle$  
($|\tilde\phi_{\mathbf{R_0},p}\rangle$) are the all-electron (pseudo) partial waves centered on the absorbing atom. The quantities $\left\langle \tilde{p}_{\mathbf{R_0},p}\right|$ are a complete set of projector functions, labeled by the index $p$. The following conditions are satisfied: 
\begin{eqnarray}
 \tilde\phi_{\mathbf{R},p}(\mathbf{R})&=&\phi_{\mathbf{R},p}(\mathbf{R}) \textrm{ outside }\Omega_\mathbf{R}\\
\left\langle\tilde{p}_{\mathbf{R},p}\middle|\tilde\phi_{\mathbf{R'},p'}\right\rangle&=&\delta_{\mathbf{R} \mathbf{R}'}\ \delta_{p p'} \textrm{ inside  }\Omega_\mathbf{R}
\end{eqnarray}
The quantity $\Omega_\mathbf{R}$ is the so-called augmentation (or core) region centered on atomic sites ${\bf R}$.
In our case, the all-electron partial waves are chosen to be the solutions of the Schr\"odinger equation 
for the isolated atom.

It is worthwhile recalling that Eq. \ref{eq:pawcrosssec2} holds under the assumption that the initial state is localized on the absorbing atom, so
that the overlap between the state $\hat{\cal O}|\psi_i\rangle$ and $ | \tilde\phi_{\bf R}\rangle$ can be neglected
if ${\bf R}\ne{\bf R}_0$. Furthermore, while Eq. \ref{eq:pawphi0} is in principle valid for a complete set of projectors, it converges after a few terms.

In the appendix section \ref{appendix} we derive the expression of the transition matrix element $\left\langle \phi_{\mathbf{R}_0,p}\middle| \hat{ {\cal O}}\middle| \psi_i \right\rangle$ for a general edge.

The single particle interpretation of x-ray absorption is no longer pertinent in the presence of a strong interaction between the photoelectron and the core hole. For the E1E1 (electric dipole) channel, this is  the case when the spin-orbit splitting of the initial state is small and the adjacent edges overlap. A notable example are the L$_{2,3}$ edges of transition elements\cite{Bunau2012} or M$_{4,5}$ edges of rare earths, no longer obeying the statistical branching ratio.\cite{Zaanen1985} At K edges, neglecting the interaction between the photoelectron and the core hole leads to the overestimation of the energy position of the E2E2 (electric quadrupole) spectral structure. \cite{Juhin2010} Often, the angular dependence of the E2E2 structure is inappropriately described in the single particle picture. \cite{Juhin2008} 

\section{Technical Details}
\label{sec:calc}

The calculation of a general edge was implemented in the XSpectra code \cite{Gougoussis2009} belonging to the Quantum Espresso distribution.\cite{QE} The method was applied to the calculation of  x-ray absorption spectra of fcc Cu (L$_{2,3}$ edges of Cu), Cu$_2$O (L$_{2,3}$ edges of Cu) and 2H-MoS$_2$ (L$_{2,3}$ edges of S).

The experimental lattice parameters are used for all three systems: $a = 3.600$ \AA$\ $ for fcc  Cu\cite{Wyckoff} (spacegroup Fm$\bar{3}$m), $a = 4.268$ \AA$\ $ for Cu$_2$O\cite{Kirfel1990} (spacegroup Pn$\bar{3}$) and, finally, $a = 3.161$ \AA$\ $ and $c = 12.295$ \AA$\ $ for 2H-MoS$_2$\cite{Schonfeld1983} (spacegroup P6$_3$/mmc).  

For Cu, Mo and O, we used ultrasoft pseudopotentials\cite{Vanderbilt} including semicore states for Cu and Mo. We use a Troullier-Martin\cite{Troullier} norm-conserving pseudopotential for S.
Three, respectively two projectors for the $l = 2$ channel, as well as one projector for the $l = 0$ channel were needed for the proper reconstruction of core states for the absorbing Cu and S atoms.
 
Cu and Cu$_2$O are described within the generalized gradient approximation (GGA) in the Perdew-Burke-Ernzerhof (PBE) parametrization.\cite{Perdew1996} In the case of 2H-MoS$_2$  we used both the local density approximation (LDA) \textit{via} the Perdew-Zunger (PZ) functional\cite{Perdew1981} and PBE, without any detectable difference.

The kinetic energy cutoff of the wavefunction expansion is chosen as $30$ Ry in Cu, $60$ Ry in Cu$_2$O, and $50$ Ry  in 2H-MoS$_2$. 
The charge density cutoff is set  ten times larger then the kinetic energy one. 

In the case of metallic Cu, integration over electron-momentum is performed by using a uniform 20 $\times$ 20 $\times$ 20
grid. In the case of insulating Cu$_2$O and 2H-MoS$_2$ we use  20 $\times$ 20 $\times$ 20 and 10 $\times$ 10 $\times$ 5  electron-momentum grids, respectively. 

The core-hole is described within a static approximation. A pseudopotential with a core-hole is generated for the absorbing atom. Furthermore, a supercell is used to minimize the interaction
between the absorbing atom and its periodic images. We use a $3 \times 3 \times 3$ supercell for Cu, a $2 \times 2 \times 2$ one for Cu$_2$O  and finally a $2 \times 2 \times 1$ supercell for 2H-MoS$_2$.

In the XANES calculations, electronic integration was performed on  a  $30 \times 30 \times 30$ grid for Cu and on a $20 \times 20 \times 10$ grid for 2H-MoS$_2$.  In the case of Cu$_2$O we use the same grid as in the self-consistent calculation.  These grids refer to the unit cell when no core-hole is involved. For supercells the grids are rescaled, due to the smaller Brillouin zone.

The spin-orbit coupling on the final (valence) states probed by the spectroscopy is neglected. Although full relativistic calculations are needed to describe heavy transition elements of the late 4$d$ and the 5$d$ series, our approximation is reasonable for the applications presented in this paper. In particular, the spin-orbit coupling acting on the 3$d$ states of transition elements subjected to crystal field is negligible.\cite{VanVleck1932,Stohr2006} However, the description of the spin-orbit coupling for the core states is mandatory in order to distinguish between related edges as the L$_2$ (transitions from 2$p_{1/2}$) and L$_3$ (2$p_{3/2}$), M$_2$ and M$_3$ or M$_4$ and M$_5$. In the appendix we show the transition matrix elements calculated accordingly. 

The radial parts of the 2$p_{1/2}$ and 2$p_{3/2}$ core wavefunctions are taken as identical, a very good approximation in the applications
treated in the present paper, as it can be explicitly verified from a DFT relativistic all electron
calculation on the isolated atom. 
We account for the spin-orbit splitting of the 2$p$ states by shifting the calculated spectra by the value of the energy separation between the L$_3$ and the L$_2$ edges.
This value can be either taken from a DFT relativistic all-electron calculation on the isolated atom or from experiments. 
The two choices are equivalent when the core-hole can be treated at a DFT level. On the contrary, strong Coulomb interaction between the 2$p$ hole and the valence $d$ electrons may lead to an  energy separation between the L$_3$ and the L$_2$ which is different from the value of the core level spin-orbit splitting.\cite{Laskowski2010}
In the absence of spin-orbit coupling on the $d$ states, the L$_2$ and L$_3$ edges have similar structures, 
while the intensities are related by a factor of 2 (the statistical branching ratio), according to the manifold of the initial states.

\begin{figure}
\includegraphics[width=0.53\textwidth]{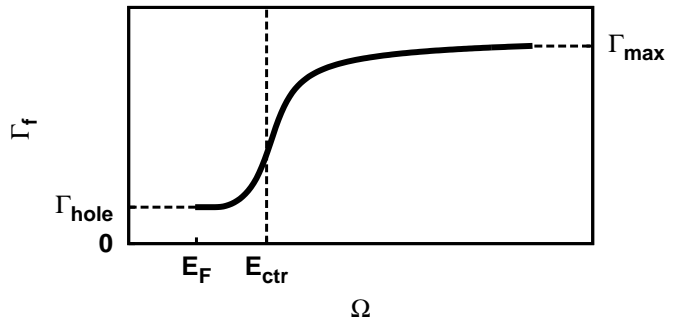}
\caption{The raw spectrum is convoluted with a lorentzian of acrtangent-like, energy-dependent width. E$_F$ is the cutoff energy, $\Gamma_{hole}$ is the spectral width at low energy, above E$_F$ and $\Gamma_{max}$ is the value at high (infinite) energy. The inflection point is situated at E$_{ctr}$.}
\label{fig:gamma_atan}
\end{figure}

The occupied states are eliminated from the XANES spectra by introducing a cutoff,  which appears implicitly in Eq. \ref{eq:sig_onebody} in the summation on the empty states $f$. The cutoff is chosen at the calculated Fermi level or at the lowest unoccupied band, depending on the metallic character of the compound. In the presence of a core-hole, the self-consistent calculation is performed on a charged cluster. This means that the energy cutoff is calculated by assuming that the photoelectron does not participate to the screening of the core-hole.

The XANES spectrum being calculated from stationary Kohn-Sham states, a broadening must be applied in order to account for the  finite life-time of the electronic levels involved in the transition: 
\begin{equation}
\sigma^{conv}(\Omega) = \int_{E_F}^{\infty} d\omega\ \sigma(\omega)\ \frac{1}{\pi}\ \frac{\Gamma_f(\Omega)}{{\Gamma_f(\Omega)}^2 + (\hbar\Omega - \hbar\omega)^2}
\label{abs_conv}
\end{equation}
The Fermi level or the lowest unoccupied band energy is labeled $E_F$.
Subsequently, the raw spectrum $\sigma(\omega)$ is convoluted with a lorentzian having an energy dependent width $\Gamma_f$:
$$\Gamma_f(\Omega) = \Big\{\begin{array}{ll} 0 & ,\hbar\Omega < E_F \\ \Gamma_{hole} + \gamma(\Omega) &, \hbar\Omega > E_F \end{array} $$
The first term  $\Gamma_{hole}$ is the core-level width, taken as energy-independent since screening is treated statically. In practice we use the standard, edge-dependent, tabulated values. \cite{Krause1979} The second term $\gamma(\Omega)$ is the spectral width due to the final state.  In general, the spectral structures at the L$_2$ edge are broader than the ones at L$_3$ due to the presence of an extra desexcitation channel, the Coster-Kr\"onig effect.\cite{Coster1935} Hence, distinct convolution widths are needed at the two edges in order to reproduce the spectral shape. Furthermore, to account for inelastic scattering events (e.g. plasmons), a Lorentzian with a more sophisticated energy dependence is sometimes preferred to a step-like function. We choose the convolution with an arctangent-like function, an empirical model close to the Seah-Dench formalism: \cite{SeahDench} 
\begin{equation}
\gamma(\Omega) =  \Gamma_{max}\left(\frac{1}{2} + \frac{1}{\pi} \arctan \left(e - \frac{1}{e^2}\right)  \right)
\label{eq:SeahDench}
\end{equation}
where $e=(\Omega-E_F)/(E_{ctr}-E_F)$. The energy dependent broadening is bound between $\Gamma_{hole}$ and $\Gamma_{max}$, the values at low and respectively, high (infinite positive) energies. The inflection point of the arctangent is situated at $E_{ctr}$. Figure \ref{fig:gamma_atan} illustrates the generic form of $\Gamma_f(\Omega)$ and the significance of the above parameters. In this work, the following values of $\Gamma_{hole}$, $\Gamma_{max}$ and $E_{ctr}-E_F$ have been used: 0.35, 4, 6 eV for Cu; 0.22, 4, 9 eV for Cu$_2$O and 0.3, 7, 6.5 eV for 2H-MoS$_2$. 

\begin{figure}
\centerline{\includegraphics[width=0.59\textwidth]{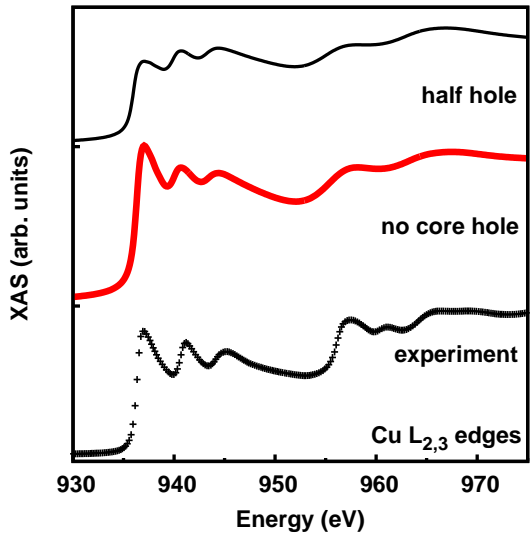}}
\caption{(color online) L$_{2,3}$ absorption edges in metallic Cu. Experimental data\cite{Cu_TEY} (dots) \textit{versus} calculations: without core-hole (thick solid) and with half core-hole (thin solid). The spectra are shifted vertically for convenience.}
\label{fig:Cu_XAS}
\end{figure}

\section{Results and discussion \label{sec:res}}

\subsection{Cu L$_{2,3}$ edges in copper}

\begin{figure}
\centerline{\includegraphics[width=0.53\textwidth]{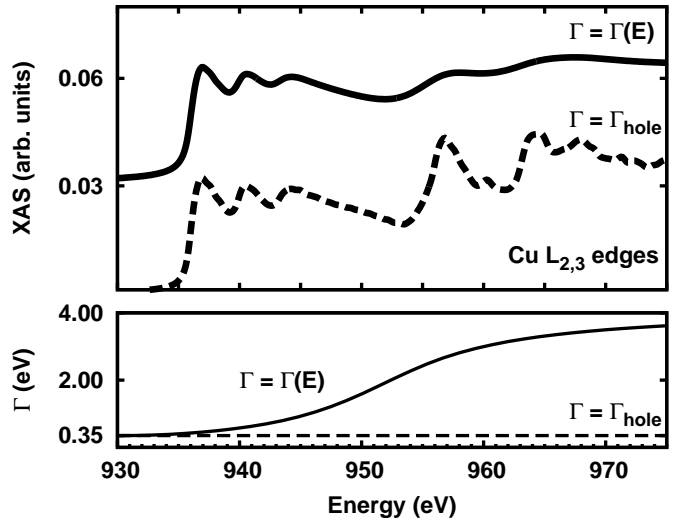}}
\caption{ Two distinct models for the convolution of spectra. Lower panel: energy independent (dashes) and arctangent-like Lorentzian-width (thin solid). Upper panel: the corresponding convoluted spectra for Cu L$_{2,3}$ absorption edges: at constant width (dashes) and arctangent-like (thick solid). }
\label{fig:Cu_XAS_gamma}
\end{figure}

To validate our implementation we first calculate L$_{2,3}$ edges for fcc Cu.
In fcc Cu, the 3$d$ states of the absorbing atom are almost entirely occupied (atomic Cu is in nominal 3$d^{10}$). 
XANES at L$_{2,3}$ edges probes the first empty $d$ states just above the main 3$d$ occupied band.
These states contribute very weakly to the overall DOS and in a narrow energy region.
For this reason, and in agreement with the result in Ref. \onlinecite{Hebert2007}, 
 we find that an extremely dense k-point sampling (electron-momentum grid) is needed
to converge the x-ray spectra (see the technical details section).

In Fig. \ref{fig:Cu_XAS} the calculated x-ray absorption (XAS) spectra are compared to experimental data. \cite{Cu_TEY} 
The effect of the core-hole at the Cu L$_{2,3}$ edges in fcc Cu was thoroughly tested.
We find that the best agreement between theory and experiment is achieved by neglecting core-hole effects. The global intensity mismatch at the L$_2$ edge (the calculated intensities are less than the measured one) is most probably due to the background subtraction in the experimental spectrum. 
The inclusion of the full core-hole (not shown) shifts the weight of the
$d$ states at several eV below the Fermi level and significantly decreases their contribution above the Fermi level.
Even with half core-hole, we find that the agreement with the experiment is worse than in the case without core-hole, with the first peak at the L$_3$ being less intense than the second one (see Fig. \ref{fig:Cu_XAS}).

Our analysis is in partial agreement with the previous interpretation of XANES data on fcc Cu in Ref. \onlinecite{Ebert1996}, based on the Korringa-Kohn-Rostoker method of band structure calculation. The only notable difference between the two calculations is that Ref. \onlinecite{Ebert1996} produces full relativistic results. In both calculations, the core-hole was not included explicitly. While the positions of the spectral structures agree in the two calculations, it is no longer the case for the intensity of the leading L$_3$ peak. More precisely, the ratio between the leading peak and  the second L$_3$ structure  in the spectrum of Ref. \onlinecite{Ebert1996} is 1.5 larger than in our calculations.

The reason for this disagreement is very likely related to the procedure adopted to eliminate the occupied states from the calculation.  We find that the intensity of the first peak in the
spectrum strongly depends on this factor,  which is often a problem in 
 metallic systems,  due to the absence of the gap in the electronic spectrum. As such, the disagreement between the two calculations
is not due to fundamental methodological differences in the calculation.

Another difference with  Ref. \onlinecite{Ebert1996} concerns the estimate of the magnitude of  transitions from 2$p$ to $s$
 states. In Ref.  \onlinecite{Ebert1996}  it was estimated to be 5\% of the total intensity, 
whereas they are completely negligible according to our results. The discrepancy may come from treatment of relativistic effects. 
\begin{figure}
\centerline{\includegraphics[width=0.53\textwidth]{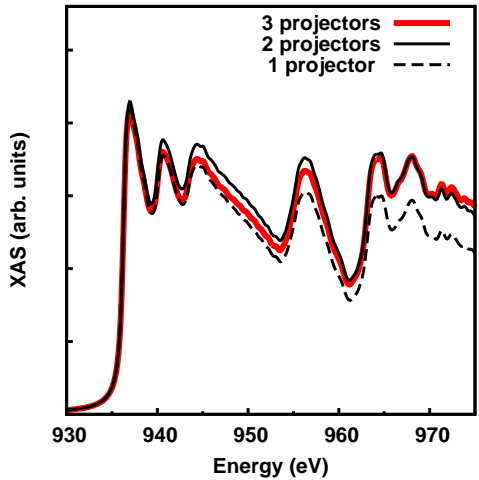}}
\caption{Dependence of the calculated  Cu L$_2$ edges on the core-level reconstruction: one (dashes), two (thin solid) and three (thick solid) projectors for the $ l = 2 $ channel.} 
\label{fig:Cu_proj}
\end{figure}

Note that the analysis of electron energy loss near edge structure spectroscopy (ELNES) data,\cite{Hebert2007,Luitz2001} carried out with a similar
formalism to the one used in XANES, leads to the conclusion that a 
partial half core-hole screening is necessary to explain the experimental data. 
 This apparent disagreement is mostly due to the differences in the experimental XANES and ELNES spectra, and not to our calculation, which actually agrees fairly well with the ones featured in Refs.  \onlinecite{Hebert2007} and \onlinecite{Luitz2001}. Indeed, the first peak L$_3$ peak in ELNES spectra\cite{Hebert2007,Luitz2001} is smaller than the second one, opposite to XANES experimental data\cite{Cu_TEY,Ebert1996} (see, for instance, Fig. \ref{fig:Cu_XAS}).

The fact that we achieve a very good description of the experimental data by completely neglecting the core-hole should be interpreted carefully. While one might be tempted to say that in metallic systems the core-hole is almost entirely screened, we recall that at K edges this is not the case. \cite{Bunau2009,Gougoussis2009} It has already been pointed out \cite{Mauchamp2009} that it is unlikely to predict whether a XANES calculation with core-hole will agree better to the experiment than the one without hole, due to the inadequacy of the description of the core-hole in the single particle approximation. Nevertheless, one can say that the 2$p$ hole is less profound than the 1$s$ one and consequently less screened. Therefore partial screening is more likely to be needed for the description of L$_{2,3}$ edges than at K edges. 

Fig. \ref{fig:Cu_XAS_gamma} features the effect of the convolution procedure on the shape of L$_{2,3}$ edge of Cu (see Eq. \ref{eq:SeahDench}). The
 inclusion of an energy dependent convolution improves the agreement with the experiment at the L$_2$ edge.
 Indeed, a Lorentzian of constant width ($\Gamma_{hole}$) overestimates the resolution at high energies, whereas the arctangent-like energy-dependent Lorentzian can reproduce the correct broadening.

We recall that the sum in Eq. \ref{eq:pawphi0} formally involves an infinite number of terms. However, experience demonstrated
 that at K edges, two linearly independent projectors for the $l = 1$ channel are enough in order to converge the E1E1 calculated XANES  
within the first $\approx$ 50 eV above the edge. \cite{Gougoussis2009} The description of the E2E2 contribution sometimes requires three projectors for the $l = 2$ channel, particularly when the energy of the first projector is set at the one of the semicore states.
It is then not obvious that two $l = 2$ projectors are sufficient for calculations of L$_{2,3}$ edges, where the sampled states have mostly $d$ character.  

Fig. \ref{fig:Cu_proj} shows the convergence of the L$_2$ spectrum with respect to the number of projectors with $d$ character considered. 
In particular, we set the first projector at the energy of the DFT all-electron 3$d$ states of atomic Cu in neutral configuration. 
The second and third projector are at $2.5$ Ry and $10.0$ Ry higher energies. This choice  leads to  linearly independent projectors. 
As shown in Fig. \ref{fig:Cu_proj}, the use of a single projector underestimate the XANES cross section already at energies just above the main L$_2$ edge. 
The use of additional projectors increases the calculated L$_2$ XANES at higher energies. 
Three projectors are by far sufficient to describe the full L$_{2}$ spectrum $50$ eV above the edge.

\begin{figure}
\centerline{\includegraphics[width=0.55\textwidth]{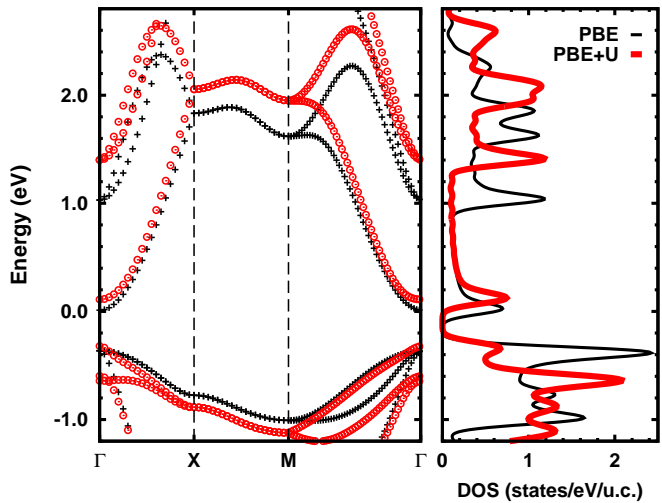}}
\caption{(color online): electronic structure of Cu$_2$O: PBE (crosses) and PBE+U (circles) with $U = 8$ eV. In the right panel, the electronic
DOS is plotted: PBE (thin solid) and PBE+U (thick solid). Positive energies are associated to empty states.}
\label{fig:Cu2O_nohole_bands}
\end{figure}

In Ref. \onlinecite{CASTEP} the authors recall that Eq. \ref{eq:pawcrosssec2} is not necessarily exact for the 2$p$ core level. Indeed, should the 2$p$ wavefunction no longer be confined within the augmentation region, Eq. \ref{eq:pawcrosssec2} is to be replaced with its more general form (see, for instance, Eq. 5 in Ref. \onlinecite{Taillefumier2002}). However, we find that in the case of Cu both 1$s$ and the 2$p$ core wavefunctions are well localized within the augmentation region, hence  the reconstruction according to Eq. \ref{eq:pawcrosssec2} is undoubtedly correct, provided that (i) a sufficient number of projectors is included and (ii) the radius of the augmentation region $\Omega_{R_0}$ is large enough to confine the 2$p$ radial function. 
  
\subsection{Cu L$_{2,3}$ edges in cuprite (Cu$_2$O)}
\label{sec:sym}

Having validated our implementation on fcc Cu, we switch to the more complex case of  Cu$_2$O. 
Cu$_2$O is a non magnetic insulator where copper is nominally in a Cu$^{+1}$, 3$d^{10}$ valence state. Density functional theory (DFT) succeeds in 
capturing the insulating ground state of Cu$_2$O, albeit with a smaller gap than the one detected in experiments. Our PBE calculations predict 
a $0.36$ eV direct gap, in agreement with previous calculations,\cite{Ruiz2003,Ching1989,Islam2009}
to be compared to the much larger experimental value, ranging from
 $2.02$ to $2.4\pm 0.3$ eV (see Ref. \onlinecite{Haidu2011} and references therein).

A way to improve the description of the gap is to include the Hubbard $U$ in the calculations, \textit{via} the DFT+U scheme. While this has been achieved in Ref. \onlinecite{Laskowski2003}, the question whether it improves the description of the empty states seen in the x-ray absorption spectroscopy is still open.

Fig. \ref{fig:Cu2O_nohole_bands} features the band structure over selected high-symmetry directions in the Brillouin zone, for $U = 8$ eV.  One can see a small opening of the gap, from $0.36$ in PBE to $0.43$ eV in PBE+U. The reason for the small gap opening has been explained in Ref. \onlinecite{Laskowski2003}, where the authors acknowledge that both DFT and DFT+U overestimate the amount of $s$-$d$ hybridization. 

While the gap increase is very small, the empty states are substantially shifted to higher energy after inclusion of the $U$ term (Fig. \ref{fig:Cu2O_nohole_bands}). This will significantly affect the shape of the calculated XANES spectra, which can be safely interpreted within the single particle picture, given the (nominally) closed shell configuration. Furthermore, by direct comparison with the experiment one can assess whether the DFT+U approximation is appropriate to the description of Cu$_2$O.

In order to see the qualitative
effect of the Hubbard term, we choose the value of  $U=8$ eV that is (i) a realistic value for Cu based oxides \cite{Gougoussis2009,Nolan2006}
and (ii) allows a substantial shift of spectral features to judge the effect of $U$ in this system.

Cu$_2$O crystallizes in a cubic structure with four equivalent Cu atoms \textit{per} unit cell. 
To calculate the absorption signal, one should consider the contribution of all  Cu atoms (one calculation for each absorber). However, 
given that the four 
Cu atoms in the unit cell are equivalent, 
the problem can be simplified by making use of symmetry considerations. Let the atomic E1E1 absorption tensor ${d}$ defined by 
$$ \sigma = \sum_i\ \sum_{\alpha\beta}\ \epsilon_\alpha\ {d}_{\alpha\beta}^{(i)}\ \epsilon_\beta$$
where $\alpha,\beta = x,y,z$ and $i=1,2,3,4$ runs over the $4$ equivalent atoms.
The matrix ${d}^{(i)}$ is hermitian and obeys the point group symmetry of the specific atom site. 
In the case of the Cu atom located at ($\frac{1}{4}$ $\frac{1}{4}$ $\frac{1}{4}$) in reduced crystal coordinates (first conventional origin choice for Pn$\bar{3}$), the corresponding tensor is (S$_{6}$ point group symmetry):
$$ {d}^{(1)} = \left(\begin{tabular}{ccc} $a$ & $b$ & $b$ \\ $b^\ast$ & $a$ & $b$  \\ $b^\ast$ & $b^\ast$ & $a$\end{tabular}\right)$$ 
with $a \neq b$ real. The remainder ${d}^{(j)}$ with $j = 2,3,4$ can be obtained by:

\begin{figure}
\centerline{\includegraphics[width=0.55\textwidth]{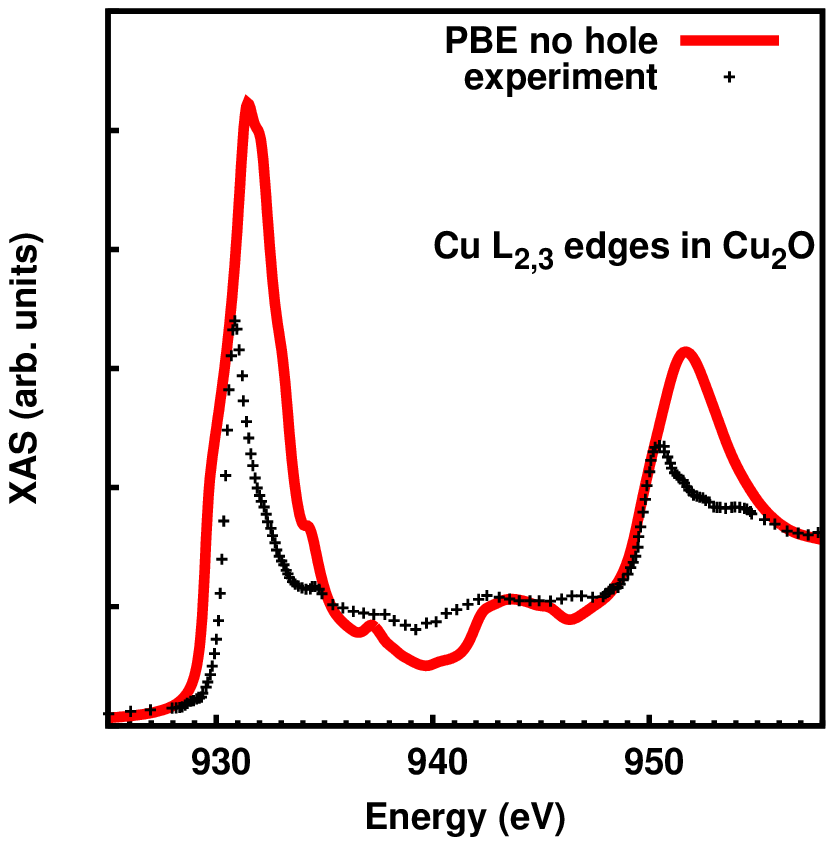}}
\centerline{\includegraphics[width=0.55\textwidth]{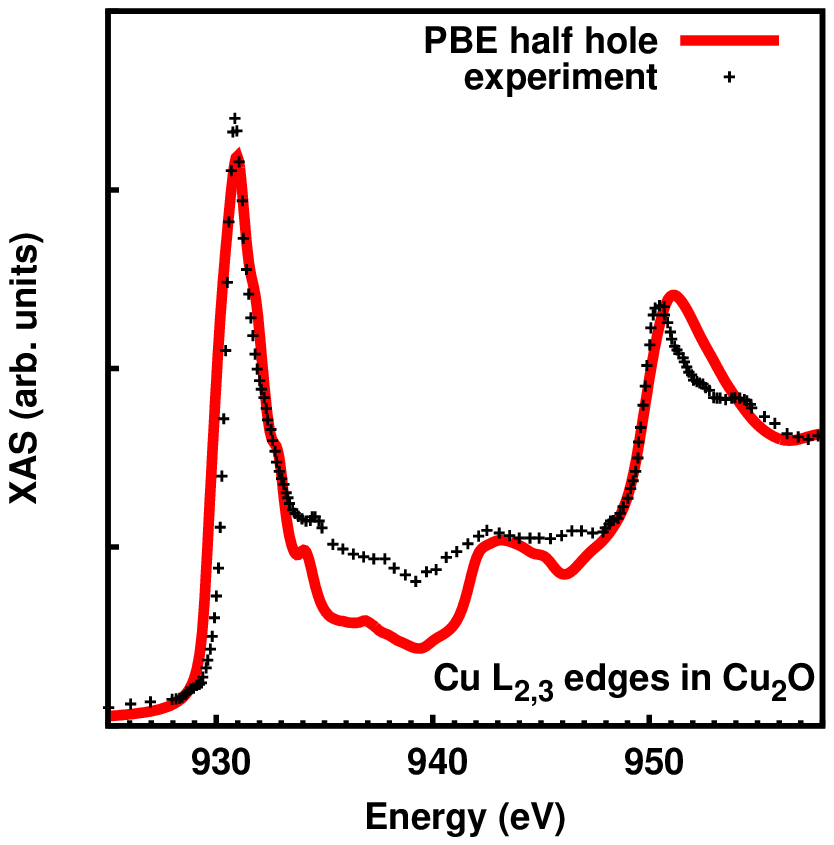}}
\caption{(color online) Calculated XANES spectra (solid) at the Cu L$_{2,3}$ edges in Cu$_2$O and comparison with experimental data from Ref. \onlinecite{Hu2008} (dots):  PBE with no core-hole (top) and with half core-hole (bottom).}
\label{fig:Cu2O_XAS_noU}
\end{figure}

\begin{figure}
\centerline{\includegraphics[width=0.55\textwidth]{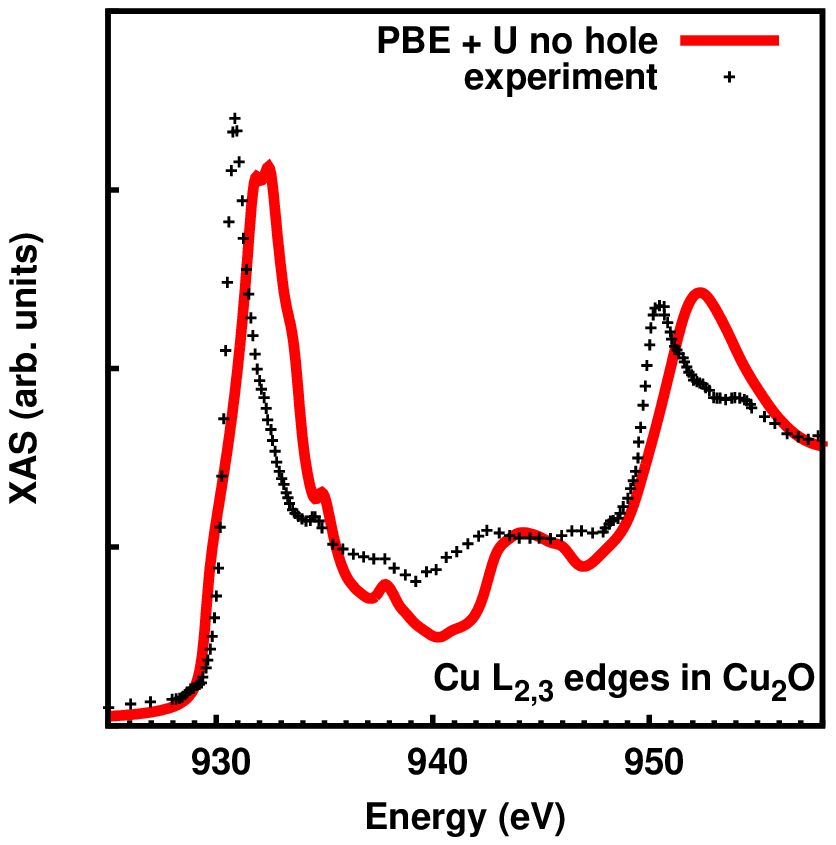}}
\centerline{\includegraphics[width=0.55\textwidth]{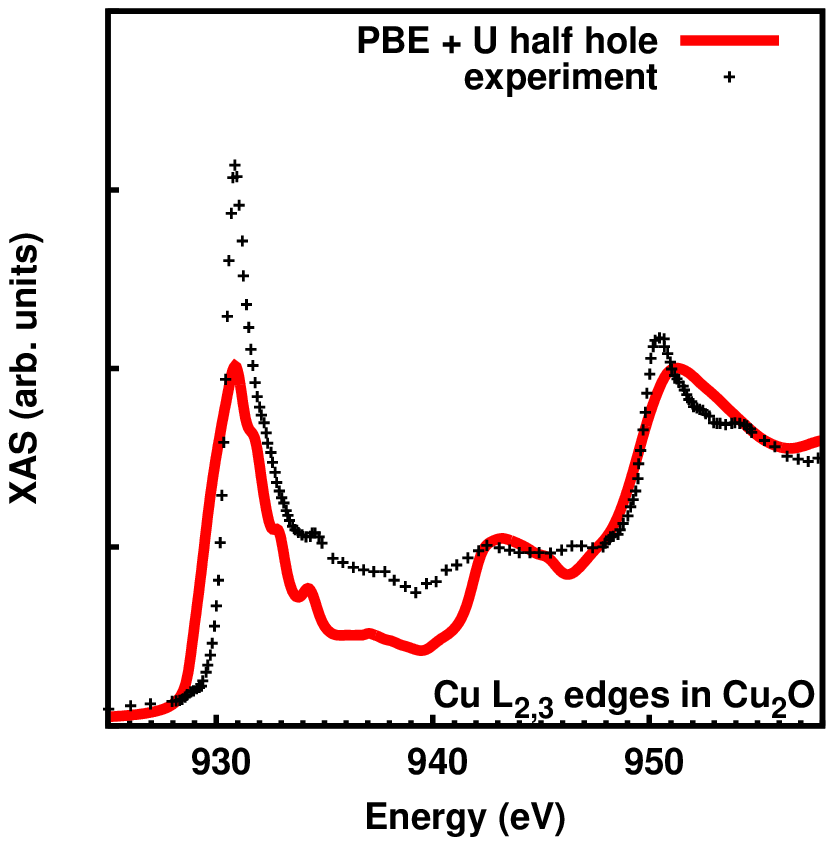}}
\caption{(color online) Calculated XANES spectra (solid) at the Cu L$_{2,3}$ edges in Cu$_2$O and comparison with experimental data from Ref. \onlinecite{Hu2008} (dots): PBE+U with no core-hole (top) and with half core-hole (bottom)}
\label{fig:Cu2O_XAS_U}
\end{figure}

$${d}^{(j)} = {\cal R}_{ij}^\dagger\ {d}^{(i)} \ {\cal R}_{ij}$$
where ${\cal R}_{ij}$ is the $3 \times 3$ cartesian rotation matrix associated to the symmetry operation relating the prototypical atom $i$ to its equivalent $j$. In our case the ($\frac{1}{4}$ $\frac{1}{4}$ $\frac{1}{4}$) Cu atom is related by $C_3$ rotations to the other Cu atoms. The crystal absorption tensor  ${\cal D}$ can be obtained as the average over all the atomic tensors:
\begin{eqnarray}
{\cal D}_{\alpha\beta} = \frac{1}{4}\sum_{i=1}^4\ {d}_{\alpha\beta}^{(i)} 
\end{eqnarray}
In matrix form:
\begin{eqnarray}
 {\cal D} = \left(\begin{tabular}{ccc} $a$ & 0 & 0 \\ 0 & $a$ & 0  \\ 0 & 0 & $a$\end{tabular}\right) 
\end{eqnarray}
Note that ${\cal D}$ is isotropic, which was to be expected for the absorption tensor of a crystal described by a cubic spacegroup. On the contrary, the atomic absorption tensors ${d}^{(i)}$ are not isotropic, a consequence of the S$_{6}$ point group symmetry. Hence we calculate the total E1E1 absorption signal \textit{per} Cu atom as the absorption of the prototypical  ($\frac{1}{4}$ $\frac{1}{4}$ $\frac{1}{4}$) Cu atom in the (100) direction. 
  
Similar to the case of Cu, we performed XANES calculations with three distinct screening schemes: 
with  core-hole, with half core-hole and without core-hole.  The results are shown, with or without the inclusion of the Hubbard $U$, 
in Figs. \ref{fig:Cu2O_XAS_U} ($U=8$ eV) and \ref{fig:Cu2O_XAS_noU} ($U=0$ eV) respectively. Calculations were normalized to the edge-jump at high energies and aligned to match the experimental peak at 934 eV. 
Similarly to the case of Cu, we find that the inclusion of a full core-hole (not shown) shifts the $d$ states completely
below the Fermi level  and hence yields unphysical results. In Fig. \ref{fig:Cu2O_XAS_U} we show that the inclusion of the Hubbard $U$ term in the description of the $d$ states of Cu in Cu$_2$O gives a very poor agreement with the experimental data, independently of the screening, contrary to what has been observed in other oxides.\cite{Juhin2010} We thus bring supplemental evidence to the finding in Ref. \onlinecite{Laskowski2003} according to which GGA+U does not contain the proper physics to describe Cu$_2$O.

The inclusion of a half core-hole is enough in order to explain the measured XANES at the L$_{2,3}$ edges of Cu. In Fig. \ref{fig:Cu2O_XAS_noU}, we show that  calculations without core-hole at $U=0$ eV overestimate the L$_3$ structure at $931$ eV. On the contrary, a good agreement with the experiments is achieved with a partial screening (half-hole). Indeed, the white line at the L$_3$ edge is well reproduced in terms of shape and intensity, as well as the positions of the $934$, $936$ and $943$ eV structures. The only remaining source of 
disagreement with the experiment is related to the underestimation of the background between the two edges.

A deeper analysis of the spectrum reveals that, as expected,  the transitions occur to the states formed by the hybridization between the e$_g$ of Cu and the $p$ states of O. Although the $s$ states of Cu and O lie in the same energy range as the empty 3$d$ states, they are not seen at the Cu L$_{2,3}$ edges. Indeed we find that transitions to states with $s$ character yield a negligible contribution to the spectrum.

\subsection{S L$_{2,3}$ edges in molybdenite (2H-MoS$_2$)} 

The 2H polytype of MoS$_2$ (2H-MoS$_2$) is a layered compound which crystallizes in a hexagonal structure. Each layer is composed of a MoS$_2$ unit and there are two layers per cell (6 atoms). Layers are bound together by weak Van der Waals interactions.  

While isotropic in the case of cubic compounds as Cu and Cu$_2$O, the E1E1 x-ray absorption signal depends on the orientation of the polarization in the case of 2H-MoS$_2$. We calculated the two independent XANES spectra corresponding to incident light polarized along the principal axis directions, in the plane and parallel to the $c$ axis, 
at the L$_{2,3}$ edges of S. Once again, two distinct sets of calculations have been performed, with and without core-hole. 

\begin{figure}
\centerline{\includegraphics[width=0.55\textwidth]{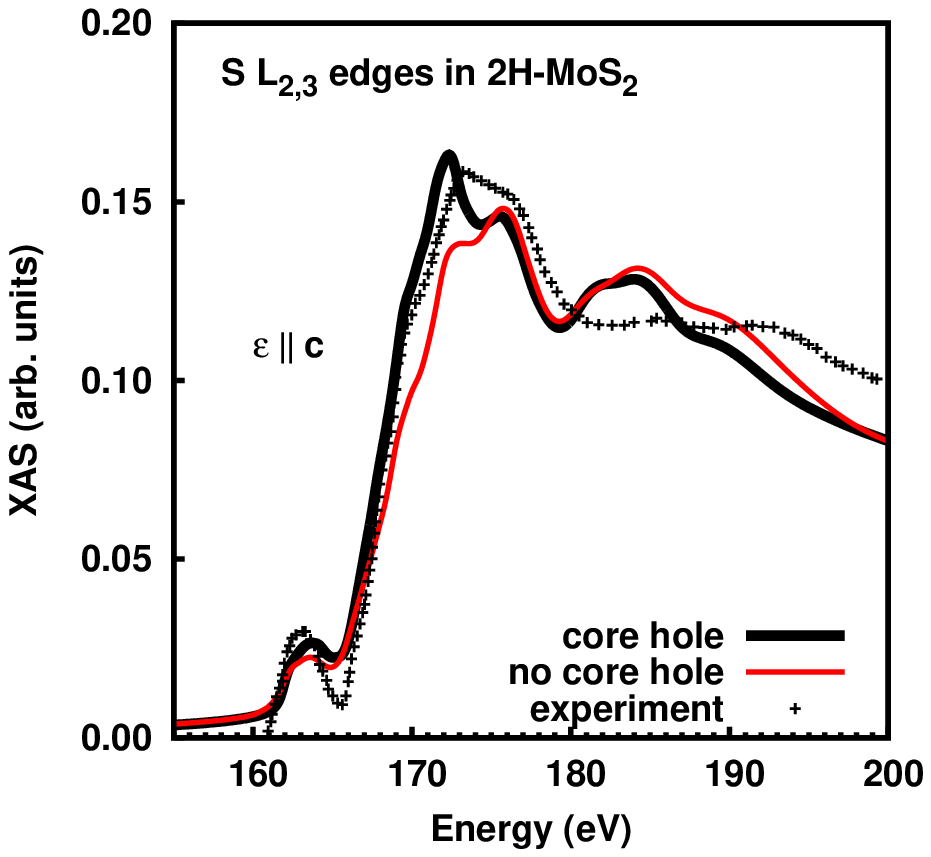}}
\centerline{\includegraphics[width=0.55\textwidth]{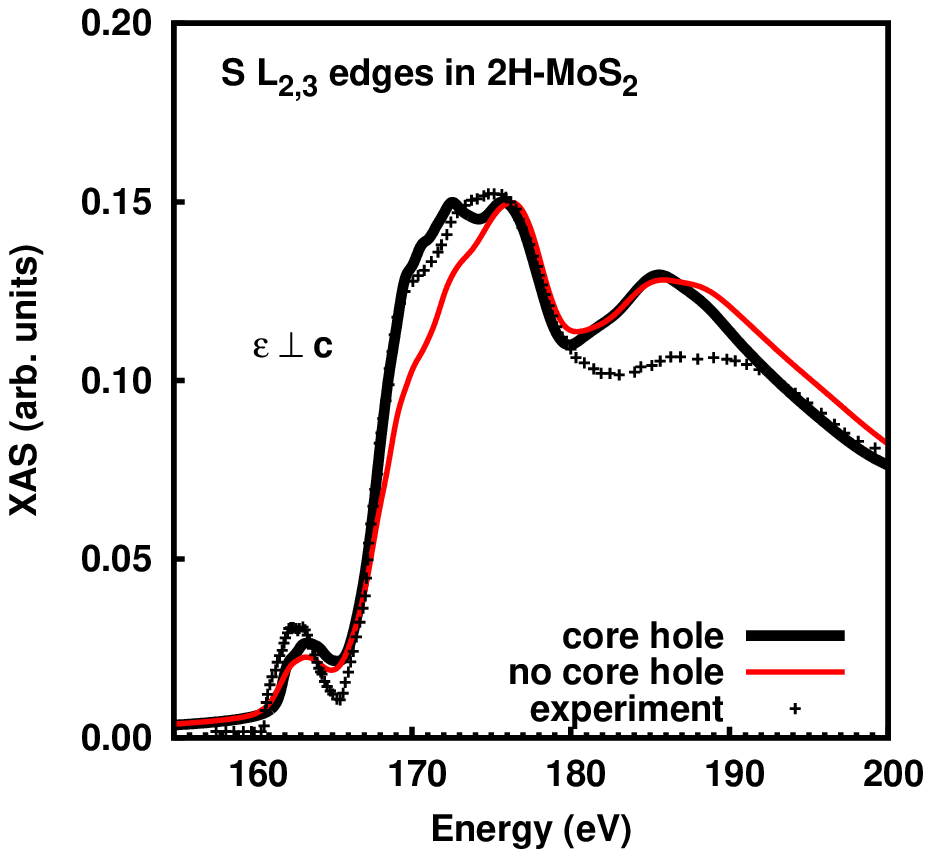}}
\caption{(color online). x-ray linear dichroism at S L$_{2,3}$ edges in 2H-MoS$_2$: calculations with (thick black solid) and without (red solid) core-hole \textit{versus} experimental data from Ref.  \onlinecite{Li1995} (black symbols). The orientation of the polarization $\hat{\bm{\varepsilon}}$ with respect to the c-axis is indicated in each panel. Calculations with core-hole were shifted with 1.5 eV to higher energies, with respect to those without hole. Calculations with and without hole were scaled independently to match the experimental data.} 
\label{fig:S_MoS2}
\end{figure}

The comparison with experimental data\cite{Li1995} for the two orientations of the incident beam polarization is shown in Fig. \ref{fig:S_MoS2}.  The experimental spectrum at $\hat{\bm{\varepsilon}}\ ||\ c$ was deduced from the two orientations published in Ref. \onlinecite{Li1995}.  Calculations including the core-hole are in better agreement with the experiment than the ones without hole, at both orientations. The agreement with the experiment is good, with the position and spectral width of peaks being reproduced at least at the main edge. We note nevertheless that the energy position of the pre-edge peak is slightly overestimated in the theoretical spectra by 1 to 1.5 eV. Second, the intensity of the pre-edge peak is underestimated by the calculation. We believe that both disagreements are a consequence of the inter-edge mixing,\cite{Bunau2012} due to the fact the S L$_{2,3}$ edges are very close in energy (1.1 eV). Third, the intensity of the spectral features above $180$ eV is significantly overestimated by the calculations, at both orientations, albeit the energy dependent convolution. We have currently no explanation for this disagreement. 

\begin{figure}
\centerline{\includegraphics[width=0.55\textwidth]{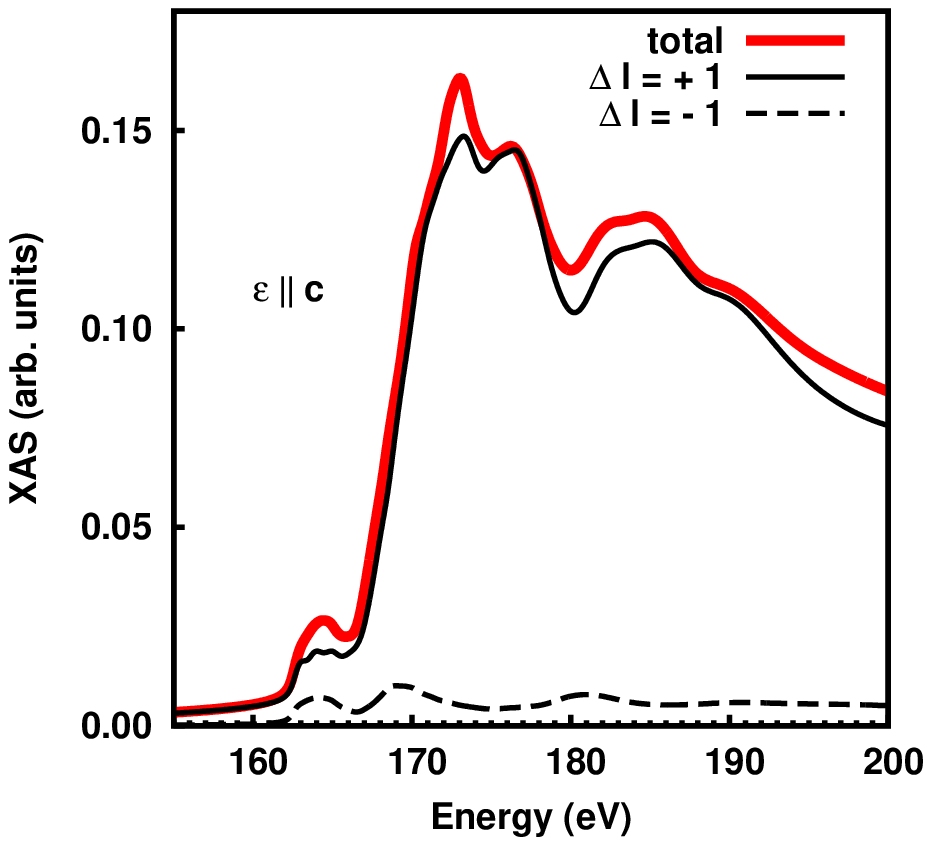}}
\centerline{\includegraphics[width=0.55\textwidth]{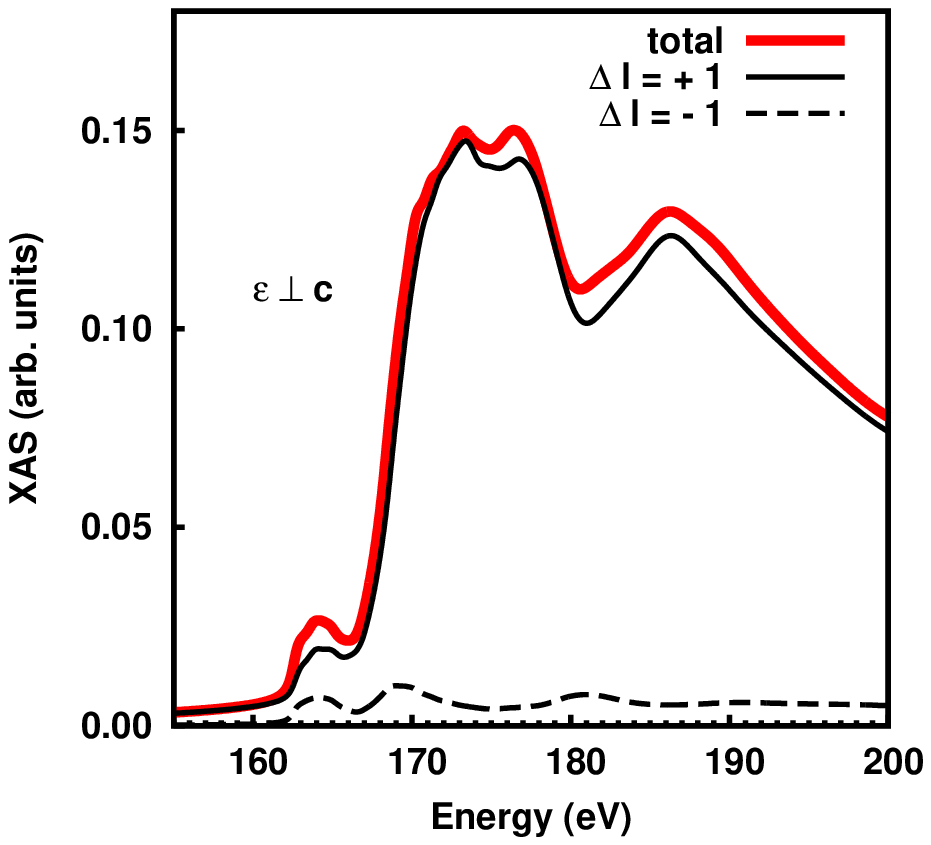}}
\caption{(color online) Decomposition of calculated XANES at the S L$_{2,3}$ edges in 2H-MoS$_2$ into angular channels. The total spectrum with core-hole (thick solid) is decomposed into transitions with $\Delta l= +1$ (thin solid) and $\Delta l= -1$ (dashes). The orientation of the polarization $\hat{\bm{\varepsilon}}$ with respect to the c-axis is indicated in each panel. The $\Delta l= -1$ channel is isotropic. The  $\Delta l= +1$ and $\Delta l= -1$ contributions do not sum up to the total spectrum, the remainder being due to cross terms.} 
\label{fig:S_MoS2_deltal}
\end{figure}

In order to reproduce the quantitative XNLD at the S L$_{2,3}$ edges in 2H-MoS$_2$, effects of both 2$p \rightarrow s$ ($\Delta l =-1$) and 2$p \rightarrow d$ ($\Delta l=+1$) transitions need to be included in the evaluation of the cross section (Fig. \ref{fig:S_MoS2_deltal}). Fig. \ref{fig:S_MoS2_deltal} shows the decomposition of the L$_{2,3}$ absorption into the two angular momenta channels, for the two polarization directions. Contrary to the common assumption that at L$_{2,3}$ edges transitions to $s$ states are negligible, Fig. \ref{fig:S_MoS2_deltal} proves they have a significant weight in a wide range of energies, albeit lesser than the one of the 2$p \rightarrow d$ transitions. This behaviour is a consequence of the significant E1 matrix element 
between the core  2$p$ and valence $s$ states. We expect it to be a general feature of L$_{2,3}$ edges of elements belonging to the third period  (Al to Ar), as suggested from weighted $s$ and $d$  DOS calculations compared with Si  L$_{2,3}$ edges in ELNES/XANES spectra of silicon\cite{Rez2008} and silicates\cite{Garvie1999}, as well as from calculations of ELNES at the L$_{2,3}$ edges of Ga in GaN. \cite{Lazar2004} In particular, transitions to $s$ states have been reported previously for XANES at the S L$_3$ edge in In$_2$S$_3$, \cite{Womes2004} based on a analysis of DOS. 

While the difference between the two sets of curves featured in Fig. \ref{fig:S_MoS2_deltal}  is due to transitions to $s$ states, one can see that the aforementioned difference depends on the polarization direction albeit the spherical character of the $s$ states. This apparent inconsistency can be understood by noting that, while the $\Delta l =-1$ transitions are indeed isotropic (Fig. \ref{fig:S_MoS2_deltal}), the total spectrum also contains cross (with respect to the selection rule), non-negligible $d$ - $s$ terms which do have an angular dependence. From Figs. \ref{fig:S_MoS2} and \ref{fig:S_MoS2_deltal} it is clear that the latter are an essential ingredient for describing the XNLD: in the absence of the cross term, the angular dependence of the 173 and 176 eV structures would nearly disappear. 

We draw the attention on a fundamental difference between $K$ and higher rank edges. At $K$ edges, the intensity of XANES structures is directly proportional to the corresponding $l$ projection of the local DOS seen in the spectroscopy - the $p$-DOS ($d$-DOS) in the case of E1E1 (E2E2) transitions. This is no longer the case at L$_{2,3}$ edges, in the presence of $d$ - $s$ cross terms, when the spectra can no longer be interpreted in terms of pure l-DOS. Note that in the isotropic limit, the $d$-$s$ interference term is strictly zero. \cite{Stohr1983} In this sense, the methodology of interpreting polarized XANES spectra in non-cubic samples by confronting them to weighted DOS is not justified. 

In this respect, a challenge for the calculation is to disentangle a particular aspect of the spectroscopy, i.e. the interference between the $\Delta l = 1$ and $\Delta l =-1$ channels, from the electronic properties of the material itself. 
One can recover information about the l-DOS from the spectroscopy calculation and group theory. To begin with, in the following argument we consider for simplicity that (i) there is no spin-orbit coupling on the 2$p$ state and (ii) the local point symmetry at the absorbing S site is described by an abelian group, with real characters. 

Defining $\langle \psi_f |=\langle s|+\langle d|$, and $|\psi_i\rangle=|p_i\rangle $ where $\langle s|$ and $\langle d|$ are
the $s$ and $d$ components of the empty valence states ($p$ being forbidden by the E1 selection rule) and $|p_i\rangle$ 
with $i=x,y,z$ is the core state, the E1 matrix element in Eq. \ref{eq:sig_onebody} is:
\begin{equation}
|\langle \psi_f | \hat {\cal D}| \psi_i\rangle|^{2}=|\langle s|\hat {\cal D} |p_i\rangle|^2+|\langle d|\hat {\cal D} |p_i\rangle|^2+
2 \langle d|\hat {\cal D} |p_i\rangle \langle p_i|\hat {\cal D} |s\rangle 
\label{eq:sd}
\end{equation}

\begin{figure}
\centerline{\includegraphics[width=0.5\textwidth]{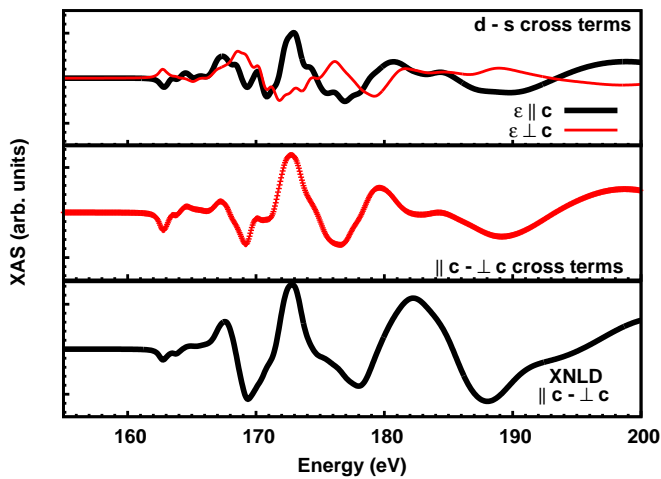}}
\caption{(color online) Assignment of the main XNLD structures to the $d$ - $s$ cross terms. Top panel: the cross terms for the two orientations, along $c$ (thick solid) and in the plane (thin solid). Middle panel: the linear dichroism of the $d$ - $s$ cross terms, taken as the difference of the curves above. Bottom panel: calculated XNLD spectrum, obtained by the subtraction of the spectra corresponding to the polarization along $c$ and perpendicular to the $c$ axis respectively. The three panels share the same scale. The comparison between the middle and bottom panels proves that the XNLD measures the anisotropy of the $d$ states, \textit{via} the mixing. The dichroic signal at 173 eV is mainly due to the $d_{z^2}$ - $s$ channel, whereas the structure at 178 eV comes from the  $d_{xy}$ - $s$  and $d_{x^2-y^2}$ - $s$  terms.}
\label{fig:S_MoS2_dic}
\end{figure}

We focus on the last term in Eq. \ref{eq:sd} that represents the $d$ - $s$  interference. It is worthwhile to recall that 
a term of the form:
\begin{eqnarray}
\langle d | r_\alpha | p_i \rangle \langle p_i | r_\alpha | s \rangle \label{eq:sym}
\end{eqnarray}
with $\alpha = x, y, z$ is zero under the above assumptions unless  (i) $\alpha$ = i and (ii) $d$ and $s$ belong to the same irreducible representation. 
This can be understood by noting that, $p_i$ and $r_i$ being in the same (one-dimensional) irreducible representation $\gamma$, 
$\gamma \times \gamma $ is the totally symmetric irreducible representation $A_1$. 
The term in (\ref{eq:sym}) is non-zero if and only if it contains the irreducible representation $A_1$, which only occurs (under the above assumptions) if  (i) and (ii) hold.  

The point group of the S site in 2H-MoS$_2$ is C$_{3v}$. In  its abelian subgroup C$_s$, $s$, on one hand,  and $d_{z^2}$, $d_{x^2-y^2}$ and $d_{xy}$ orbitals on the other hand, belong to the same irreducible representation and, as argued in the previous paragraph, can mix. One can easily see that in the $\hat{\bm{\varepsilon}}\ ||\ c$ direction the cross term is $d_{z^2}$ - $s$, while for the $\hat{\bm{\varepsilon}} \perp c$ orientation the cross term is due to $d_{x^2-y^2}$ - $s$ and $d_{xy}$ - $s$ contributions. This conclusions remains valid if one considers the spin-orbit coupling on the 2$p$ level. Under these circumstances, while the $d_{yz}$ - $s$ and $d_{xz}$ - $s$ contributions are not explicitly forbidden by symmetry, their contribution cancels after summation on the initial states belonging to the 2$p_{1/2}$ or 2$p_{1/2}$ manifold.

The difference between the $d$ - $s$ interference in the two directions (the linear dichroism) is related to the anisotropy of the $d$ orbitals. More specifically, the linear dichroism of the $d$ - $s$ channel  probes the difference between the $d_{z^2}$ - $s$ term, on one hand, and the joint contribution of $d_{x^2-y^2}$ - $s$ and $d_{xy}$ - $s$, on the other hand. This conclusion is equally valid had we considered the true C$_{3v}$ symmetry. 
 
We have already seen in Fig. \ref{fig:S_MoS2_deltal} that the main dichroic effects at 173 and 176 eV are due to the angular dependence of the cross terms. In Fig. \ref{fig:S_MoS2_dic} we continue the analysis by comparing the calculated XNLD (bottom panel), taken as the difference between the $\hat{\bm{\varepsilon}}\ ||\ c$ and $\hat{\bm{\varepsilon}} \perp c$ spectra, to the cross $d$  - $s$ term in the two orientations (top panel) and its linear dichroism (middle panel). The three insets sharing the same scale, it becomes clear that the XNLD can be safely interpreted in terms of $d$ - $s$ interference. Surprisingly, the pure $\Delta l = 1$ channel has no clear signature in the XNLD, except around 183 eV. While we have shown that the linear dichroism of the cross terms measures the $d$ states anisotropy in 2H-MoS$_2$, according to Fig. \ref{fig:S_MoS2_dic} (middle and bottom panels) this is also the case for the XNLD spectra. One can therefore assign the dichroic feature at 173 eV to $d_{z^2}$ - $s$   interference, while the feature at 176 eV is attributed to $d_{x^2-y^2}$ - $s$ and  $d_{xy}$ - $s$ mixing. In spite of the peculiar interference effect, XNLD still measures in-plane / out-of-plane $d$ anisotropy. 

On the basis of the angular dependence, the authors of  Ref. \onlinecite{Li1995} assign the pre-edge peak and the shoulder at 170 eV to transitions to $s$ states. The decomposition in Fig. \ref{fig:S_MoS2_deltal} contradicts this conclusion: while there exists a contribution of the $s$ states, it is definitely not dominating. A careful analysis of the DOS (not shown) shows that at the energies corresponding to the pre-edge peak the $s$ states of S hybridize with the $d$ orbitals of Mo. Altogether this proves the importance of calculations in order to accurately interpret the spectroscopic data. 

We recall that the interpretation of L$_{2,3}$ x-ray absorption spectra in terms of single particle $s$ and $d$ orbitals is not exact in the presence of inter-edge mixing. However, we do not expect it to dramatically affect the main edge, where the probed $d$ states are rather delocalized. This, together with the fact that the agreement with the experimental data is satisfactory, re-confirms the choice of our methodology. 

The $d$ - $s$ cross term has also been depicted at the $L_{2,3}$ edges of Ag in magnetic multilayers,\cite{Jaouen2008} albeit not in the XANES spectra but in the x-ray magnetic circular dichroism (XMCD). This suggests that a correct description of the cross term is equally important for magnetic spectroscopy. The spectroscopic interpretation would be even richer in the magnetic case, due to the exchange splitting within the 2p$_{1/2}$ and 2p$_{3/2}$ manifold. 

\section{Conclusions}

In this work we have developed a technique based on DFT and on the PAW formalism in order to obtain the x-ray absorption cross
section at a general edge, in both E1 and E2 approximations. We applied the method to the calculation of Cu  L$_{2,3}$ edges in fcc 
Cu and Cu$_2$O, as well as to S  L$_{2,3}$ edges in 2H-MoS$_2$. 
In metallic Cu we find a good agreement with experimental data 
without having to include a core-hole. 
On the contrary core-hole effects are relevant in  insulating Cu$_2$O and 2H-MoS$_2$, where their inclusion is essential in order to obtain a good description of the experimental data.

In the case of Cu L$_{2,3}$ edges in Cu$_2$O we equally study the effect of
the Hubbard $U$ term on the x-ray absorption spectra. In contrast to results on other oxides,  we find that inclusion of $U$ worsens the agreement 
with experimental data. 

Finally in the case of S L$_{2,3}$ edges in 2H-MoS$_2$,  we show that
 transitions to $s$ states yield a non-negligible contribution to the  
 S L$_{2,3}$ spectra in 2H-MoS$_2$, in a wide range of energies. 
 We believe that this is a general
property of L$_{2,3}$ edges of third row elements from Al to Ar. 
We equally point out that when cross $d$ - $s$ terms are significant, the L$_{2,3}$ spectra
are no longer to be interpreted in terms of projections of the DOS. The mixing is essential as
to interpret the XNLD at S L$_{2,3}$ edges in 2H-MoS$_2$. In such case it is utmost important to
disentangle the two channels, which can only be achieved \textit{via} first principles calculations. 

We have proven that single-particle approaches can be very useful for the interpretation of L$_{2,3}$ edges when the inter-edge mixing is negligible.
In this case \textit{ab initio} XANES calculation can assign spectral structures to various angular moment channels, and explore their possible interference.
\appendix
\section{The E1 and E2 transition matrix elements\label{appendix}}

For completeness we provide the full analytical expressions of the transition matrix element
$\langle \phi_{\mathbf{R}p}| \hat{\cal O}| \psi_i \rangle $, at a general edge, as it was implemented in the XSpectra module of the Quantum Espresso distribution.

By considering the spin-orbit coupling, the initial state (the core-level) $ | \psi_i \rangle $ is written as: 
\begin{eqnarray}
\psi_{n_il_ijm_j}({\bf r}) &=& \sum_{m_i=-l_i}^{l_i}\ \sum_{m_s = -s}^{s} \langle l_im_i, sm_s | jm_j \rangle \nonumber \\
&\times & R_{n_il_im_s}(r)Y_l^{m_i}(\hat r)\chi_{m_s}^s\nonumber \\
\end{eqnarray}
where $n_i$, $l_i$, $m_i$, $s = 1/2$ and $m_s$ are the usual notations for the quantum numbers associated to the initial (core) state in the uncoupled basis and $ \langle l_im_i, sm_s | jm_j \rangle$ the Clebsch-Gordan coefficients, whose tabulated values can be found in Ref. \onlinecite{Brouder1990}. The functions
 $R_{n_il_im_s}(r)$ and $Y_l^m (\hat r)$ are the radial and angular
 (complex spherical harmonics) parts, respectively, of the
atomic wavefunction. $\chi_{m_s}^s$ is the $m_s$ component of the spinor of spin $s$. 
The total momentum quantum numbers  $j=l_i-s, ...l_i+s$ and $m_j = -j, ..., j$ define the core-level in the $\bm{l}\cdot\bm{s}$ coupled basis. 
The Clebsch-Gordan coefficients are zero unless the following holds:
\begin{eqnarray}
m_j = m_s + m_i
\end{eqnarray}

The all-electron partial waves $\phi_{{\bf R}p}({\bf r})$ are chosen as solutions of the Schr\"odinger equation of the
isolated atom. As such they are written as 
\begin{eqnarray}
\phi_{{\bf R}p}({\bf r})   = \sum_{lm}\ \sum_\sigma\ R_{pl\sigma}(r)\ Y_l^m(\hat r)\ \chi_{\sigma}^s
\end{eqnarray}
where $p$ labels the radial wavefunction at a given energy and $n,l,m,s,\sigma$ are angular and spin quantum numbers 
associated to the partial wave. 

The electric dipole transition operator can be expanded in the spherical harmonics basis as
\begin{eqnarray}
{\cal D}=\hat{\bm{\varepsilon}}\cdot\bm r = \frac{4\pi}{3}\ r\ \sum_{\mu=-1}^{1}\ (-1)^\mu\ Y_1^\mu(\hat r) Y_1^{-\mu}(\hat \varepsilon)
\end{eqnarray}
with the E1 transition matrix element becoming:
\begin{eqnarray}
\langle \phi_{\mathbf{R} p}| &{\cal D}& | \psi_{n_il_ijm_j} \rangle = \nonumber \\
&\times&\ \frac{4\pi}{3}{\sum_{m_i = -l_i}^{l_i} \sum_{m_s = -s}^{s}} \delta_{m_j,m_i+m_s} \langle l_i m_j-\sigma, s\sigma | jm_j \rangle\nonumber \\
&\times&\ \sum_{\mu=-1}^{1}\ (-1)^\mu \sum_\sigma\ \sum_{lm}\ Y_1^{-\mu}(\hat \varepsilon)\ \chi_{m_s}^s\ \chi_{\sigma}^{s\ast} \nonumber\\
&\times&\ \int\ d\hat r\ Y_{l_i}^{m_i}(\hat r)\ Y_1^\mu(\hat r)\ Y_l^{m \ast}(\hat r)\ \nonumber \\
&\times&\ \int\ dr\ r^3\ R_{n_il_im_s}(r)\ R_{pl\sigma}(r)
\end{eqnarray}
By using the orthogonality of the spin functions:
\begin{eqnarray}
\sum_{m_s}\sum_{\sigma}\ \chi_{m_s}^s\ \chi_{\sigma}^{s\ \ast} &=& \sum_{m_s}\sum_{\sigma}\ \delta_{\sigma m_s}\nonumber \\
\end{eqnarray}
and the expression for the Gaunt coefficients (the integral is over the solid angle):
\begin{eqnarray}
{\cal G}_{l_im_i,1\mu}^{lm} = \int\ d\hat r \ Y_l^{m\ \ast}(\hat r)\ Y_{l_i}^{m_i}(\hat r)\ Y_1^\mu(\hat r)\nonumber \\
\end{eqnarray}
the E1 transition matrix element finally gives: 
\begin{eqnarray}
\langle \phi_{\mathbf{R} p}|& {\cal D} &| \psi_{n_il_ijm_j} \rangle  =\frac{4\pi}{3}\ \sum_\sigma\ 
 \langle l_i m_j-\sigma, s\sigma | jm_j \rangle \nonumber \\
&\times&\sum_{\mu=-1}^{1}\ (-1)^\mu\  Y_1^{-\mu}(\hat \varepsilon)\ \sum_{lm}\ {\cal G}_{l_i\ m_j - \sigma, 1\mu}^{lm}\nonumber \\ 
&\times& \int_{\Omega_R}\ dr\ r^3\ R_{n_il_i\sigma}(r)\ R_{pl\sigma}(r) \
\end{eqnarray}
where $\Omega_R$ is the augmentation region.

The electric quadrupole transition operator can be written as:
\begin{eqnarray}
 {\cal Q}&=& \frac{i}{2}(\hat{\bm{\varepsilon}} \cdot \bm r) (\bm k \cdot \bm r) =\nonumber \\ 
&=&\frac{i}{2}k\left( \frac{4\pi}{3}\ r\right)^2\ 
\sum_{\mu=-1}^{1}\ Y_1^\mu(\hat r) Y_1^{\mu\ast}(\hat \varepsilon)
 \sum_{\lambda=-1}^{1}\ Y_1^{\lambda\ast}(\hat r) Y_1^{\lambda}(\hat k)\nonumber \\
\end{eqnarray}
so that the E2 transition matrix element is
\begin{eqnarray}
\langle \phi_{\mathbf{R} p}|& {\cal Q} &| \psi_{n_il_ijm_j} \rangle  = \nonumber \\
&\times& \frac{i}{2}\left( \frac{4\pi}{3} \right)^2
\sum_\mu\ \sum_\lambda \ Y_1^{\mu\ast}(\hat \varepsilon)\ Y_1^\lambda(\hat k)\nonumber \\
&\times&  \sum_{lm\sigma}\ 
  \langle l_i m_j-\sigma, s\sigma | jm_j \rangle \ Y_l^m(\hat r)\  {G}_{lm, 1\lambda}^{l_i\ m_j - \sigma,1\mu}\nonumber\\
&\times& \int_0^R\ dr\ r^4\ R_{n_il_i\sigma}(r)\ R_{pl\sigma}(r) 
\end{eqnarray}
The generalized Gaunt coefficients are defined as:
\begin{eqnarray}
&&{G}_{l_1m_1,l_3m_3}^{l_2m_2,l_4m_4} =\nonumber \\
&&= \int\ d\Omega\ Y_{l_1}^{m_1\ast}(\Omega)\ Y_{l_2}^{m_2}(\Omega)\ Y_{l_3}^{m_3\ast}(\Omega)\ Y_{l_4}^{m_4}(\Omega) 
\nonumber \\
\end{eqnarray}
and the following relation holds:
\begin{eqnarray}
{G}^{l_1m_1,l_3m_3}_{l_2m_2,l_4m_4} =\sum_{lm} {\cal G}_{l_2m_2,l_1m_1}^{lm}\ {\cal G}_{l_3m_3,l_4m_4}^{lm}
\end{eqnarray}

Note that to calculate the cross section in Eq. \ref{eq:pawcrosssec2}, the contribution from each initial state $\psi_i$ from the 2$p_{1/2}$ (2$p_{3/2}$) must be evaluated independently. Initial states with distinct $jm_j$ quantum numbers do not cross in the single particle picture.

\begin{acknowledgments}
We acknowledge J. C. Cezar and H. Tolentino for providing the low
temperature L$_{2,3}$ edges XANES data for fcc copper, taken at the ID08 beamline at the ESRF. 
We acknowledge discussions with L. Paulatto, Ph. Sainctavit,
D. Cabaret, C. Brouder, G. Radtke and Y. Joly. 
We thank D. Cabaret and C. Brouder for a careful reading of the manuscript. 
One of the authors (O.B.) declares non-competing financial support
from the French ANR-BLANC-SIMI7-2010-0 program 
(SWITCH project) and equally from MINECO MAT2011 / 23791 and Aragonese IMANA projects, partially funded by the FEDER program and the European Social Fund. 
Calculations were performed at the IDRIS supercomputing center (project number 2012-100172).
\end{acknowledgments}

%

\begin{thebibliography}{10}
\bibitem{Cu_TEY}%
  \BibitemOpen
  \bibinfo {note} {Courtesy of J.C. Cesar and H. Tolentino. Obtained at the
  ID08 beamline at the ESRF, in the total electron yield detection mode and
  with linear polarization in a single crystal of Cu whose surface plane was
  oriented along the (100) direction. The dark current was measured by the
  yield in a gold mesh.}%
  \bibAnnoteFile{Stop}{Cu_TEY}%
\end{thebibliography}
\end{document}